\documentstyle[12pt,psfig]{article}
\begin{document}
\renewcommand{\thefootnote}{\fnsymbol{footnote}}

\title{Renormalization-Group Theory and Universality\\
along the Lambda Line of $^4$He}

\author {V. Dohm\\
Institut f\"ur Theoretische Physik, Technische Hochschule Aachen\\
D-52056 Aachen, Germany}
\date{}
\maketitle
\begin{abstract}
The present status of the renormalization-group (RG) predictions on the 
superfluid density $\rho_s$ and the specific heat $C^+$ and $C^-$
near the $\lambda$-line of $^4$He is briefly reviewed. Particular attention is 
given to universal amplitude ratios related to these quantities.
The goals of a new theory project are presented that involves higher-order
calculations of the amplitude functions of $\rho_s, C^+$ and $C^-$
which are of fundamental importance for a test of the universality predictions
of the RG theory.
\end{abstract}

\section{Introduction}

One of the fundamental achievements of the renormalization-group (RG)
theory of critical phenomena is the identification of universality
classes in terms of the dimensionality $d$ of the system
and the number $n$ of components of the order parameter \cite{1}.
Specifically, RG theory predicts that, within a given universality class,
the critical exponents, certain amplitude ratios and scaling functions are 
universal quantities that do not depend, e.g., on the strength of the interaction 
or on thermodynamic variables (such as the pressure).
The superfluid transition of $^4$He belongs to the $d=3, n=2$
universality class and provides a unique opportunity for an experimental
test of the universality prediction by means of measurements of the 
critical behavior at various pressures $P$ along the $\lambda$-line
$T_\lambda (P)$ (and along the $\lambda$-line of $^3$He $-$ $^4$He mixtures). 
Early tests have been performed by Ahlers and collaborators and consistency
with the universality prediction was found within the experimental resolution
\cite{2}. At a significantly higher level of accuracy, 
the superfluid density $\rho_s$ and
the specific heat $C^+$ and $C^-$ (or, equivalently, thermal expansion
coefficient $\beta^{\pm}$) above and below $T_\lambda (P)$ are planned
to be measured in the Superfluid Universality Experiment (SUE) \cite{3}
under microgravity conditions or at reduced gravity in the low-gravity 
simulator \cite{4}. 
As demonstrated recently \cite{5}, this would allow to perform 
measurements  up to $|t| \simeq 10^{-9}$ 
in the reduced temperature $t = (T-T_\lambda (P))/T_\lambda (P)$.\\
\\

Important steps towards an improved universality test can already be
performed, even in the presence of gravity, by new  ground-based 
measurements provided that
non-universal and universal effects are properly separated in a nonlinear RG
analysis of the data \cite{6}. To extract the leading 
critical exponents of $\rho_s$
and $C^\pm$ from the experimental data and to demonstrate
their universality at a highly quantitative level requires detailed
knowledge on certain universal amplitude ratios and on non-asymptotic 
corrections to the asymptotic power laws. These quantities can be calculated
by means of field-theoretic RG methods applied to the $\varphi^4$ model 
\cite{1,6}.\\ 
\\

In the following we point to a
serious lack of quantitative knowledge in the theoretical literature \cite{1}
on these amplitude ratios and correction terms. We also discuss the
implications of very recent
two-loop results for $\rho_s$ \cite{7} and shall summarize
the goals of a new theory project \cite{8} on higher-order RG calculations 
of the amplitude functions
of $\rho_s$ and $C^\pm$.

\section{Universal amplitude ratios}

In the previous analyses of experimental data for $\rho_s$ and
$C^{\pm}$ (or $\beta^{\pm}$) the following representations were employed
\cite{2,5,9}

\begin{equation}
\rho_s \quad = \quad 
k_B T_\lambda (m/\hbar)^2  A_\rho (1\; + \; k_1 \; |t| \;) 
|t|^{\zeta}
\; (1\; + \; a_\rho |t|^{\Delta} )\qquad , 
\end{equation}
\\
\begin{equation}
C^+ \quad = \quad \frac{A^+}{\alpha} |t|^{-\alpha} 
( 1\; + \; {a_c^+} |t|^{\Delta}
\; + \; E^+ |t| ) \quad + \quad B \qquad , \qquad t > 0
\end{equation}
\\
\begin{equation}
C^- \quad = \quad \frac{A^-}{\alpha} |t|^{-\alpha} 
( 1 \; + \; {a_c^-} |t|^{\Delta}
\; + \; E^- |t| ) \quad + \quad B \qquad , \qquad t < 0
\end{equation}
\\
where $m$ is the helium mass.
These representations contain 10 nonuniversal amplitudes 
$A_\rho, k_1, a_\rho, A^{\pm}, {a_c^\pm}, E^\pm, B$ which need to be
adjusted at each pressure. Their values vary by $15-30 \%$, typically, along
the $\lambda$-line. The exponents $\alpha$ and 
$\zeta = \nu (d-2) =  (2-\alpha)(d-2)/d$ and
the Wegner exponent $\Delta$ are predicted to be universal. 
The existing theoretical predictions on these exponents \cite{10} are based
on field-theoretic calculations to five-loop order \cite{11} and Borel
resummation. The recent experimental result \cite{5} $\alpha = -0.01285 \pm
0.00038$ is consistent with but more accurate than the RG estimate \cite{10}.
In addition, RG theory predicts the amplitude ratios 
\\
\begin{equation}
A^+/A^-,\qquad (A^-)^{1/d}/A_\rho, \qquad  {a_c^-}/a_\rho,
\qquad {a^+_c}/{a^-_c}\qquad 
\end{equation}
\\
to be universal \cite{1}.  Knowing their universal values for the $d=3$, 
$n=2$ universality
class would impose significant constraints on the analysis of the 
experimental data and would thereby improve the reliability
and precision of the
experimental results regarding the universality of $\alpha$ and $\zeta$.\\
\\

The present theoretical knowledge of the ratios (4), however,
is based only on low-order (mainly 1- and 2-loop) field-theoretic 
calculations
\cite{1} which imply an uncertainty at the level of at least $10 - 30 \%$. 
For example, even the
sign of the ratio ${a^-_c}/a_\rho$ was incorrectly predicted by the
two-loop $\varepsilon(=4-d)$ expansion \cite{12}. This issue was resolved by
a (low-order) calculation within the $d=3$
field theory \cite{13}. The present status of the experimental and theoretical
values of the universal ratios (4) is summarized  in Table 1 
(without error estimates).

\vspace{1.5cm}
{\bf{Table 1}} Universal amplitude ratios for the $n = 2, d = 3$
universality class
\vspace{1.5cm}

\begin{tabular}{l|c|c|c|c}  \\ \hline
       &           &                       &                  &\\
       & $A^+/A^-$ &  $(A^-)^{1/3}/A_\rho$ & ${a^-_c}/a_\rho$ & $a^+_c/a^-_c$\\ 
       &           &                       &                  &   \\ \hline
       &           &                       &                  &   \\
Experiment \cite{2}& 1.067     &  0.85 & $-$0.068 & 1.03\\
      &           &             &    & \\ \hline 
      &           &             &    & \\
Experiment \cite{14}               & 1.088  & & & 0.85\\
     &            &             &    & \\ \hline
     &            &             &    &  \\
$d=3$ Field Theory \cite{13}       & 1.05\quad & 0.78 & $-$0.045 & 1.6\\
     &  &   &  & \\ \hline
     &  &   &  &  \\
$\varepsilon-$Expansion \cite{1}& 1.029 & 1.0 & 1/6 & 1.17 \\
     &  &   &  &  \\ 
\end{tabular}
\vspace{2cm}

As noted in Ref. 13, there is a large uncertainty of the theoretical
value for ${a^+_c}/{a^-_c}$. The $d=3$ field-theoretical values for 
$(A^-)^{1/3}/A_\rho$
and ${a^-_c}/{a_\rho}$  are based on one-loop results \cite{13}.
Very recent $d=3$ two-loop results \cite{7}, to be discussed in the
subsequent section, imply considerably different values \cite{15}, 
thus indicating a considerable lack of reliability of low-order 
results for the
amplitude ratios (4).
\newpage
\section {Recent results on $\rho_s$ in two-loop order}

In an effort to improve the theoretical prediction on these amplitude
ratios, a two-loop calculation of $\rho_s$ was recently carried out by
Burnett, Str\"osser and Dohm within the $\varphi^4$ field theory in
$d=3$ dimensions \cite{7}.
The superfluid density can be calculated via Josephson's definition \cite{16}\\
\\
\begin{equation}
\rho_s \; = \;
k_B T (m/\hbar)^2 |\psi_0|^2  \frac{\partial}{\partial k^2} 
\chi_T (k)^{-1}/_{k=0}
\end{equation}
\\
where $\chi_T(k)$ is the transverse susceptibility at finite wave number $k$
and $\psi_0$ is the order parameter (Bose condensate wave function). 
Within the $d=3$ $\varphi^4$ field theory the representation (5)
becomes \cite{7,13}
\begin{equation}
\rho_s \; = \; k_B T (m/\hbar)^2 {\xi^{-1}_-} f_\psi (u) f_T(u)
\end{equation}
\\
where  
\\
\begin{equation}
\xi_- \; = \; \xi^-_0 |t|^{-\nu} (1 + a^-_\xi |t|^\Delta 
+ \cdot \cdot \cdot)
\end{equation}
\\
is an appropriately defined correlation length.  
The amplitude functions $f_\psi(u)$ and $f_T(u)$ depend on the effective 
renormalized $\varphi^4$ coupling $u = u({\xi^{-1}_-}$) which for 
$\xi_- \rightarrow \infty$ approaches the fixed point value 
$u(0) = u^* = 0.0362$ \cite{17}.  These functions
are plotted in Fig. 1 in zero-, one-, and two-loop order. Their 
fixed point values $f_{\psi}(u^*)$ and $f_T(u^*)$ determine the asymptotic
amplitude $A_\rho$ in (1) while their derivatives at $u = u^*$ contribute to
the subleading amplitude $a_\rho$.

      \begin{center}
        \leavevmode
        \psfig{figure=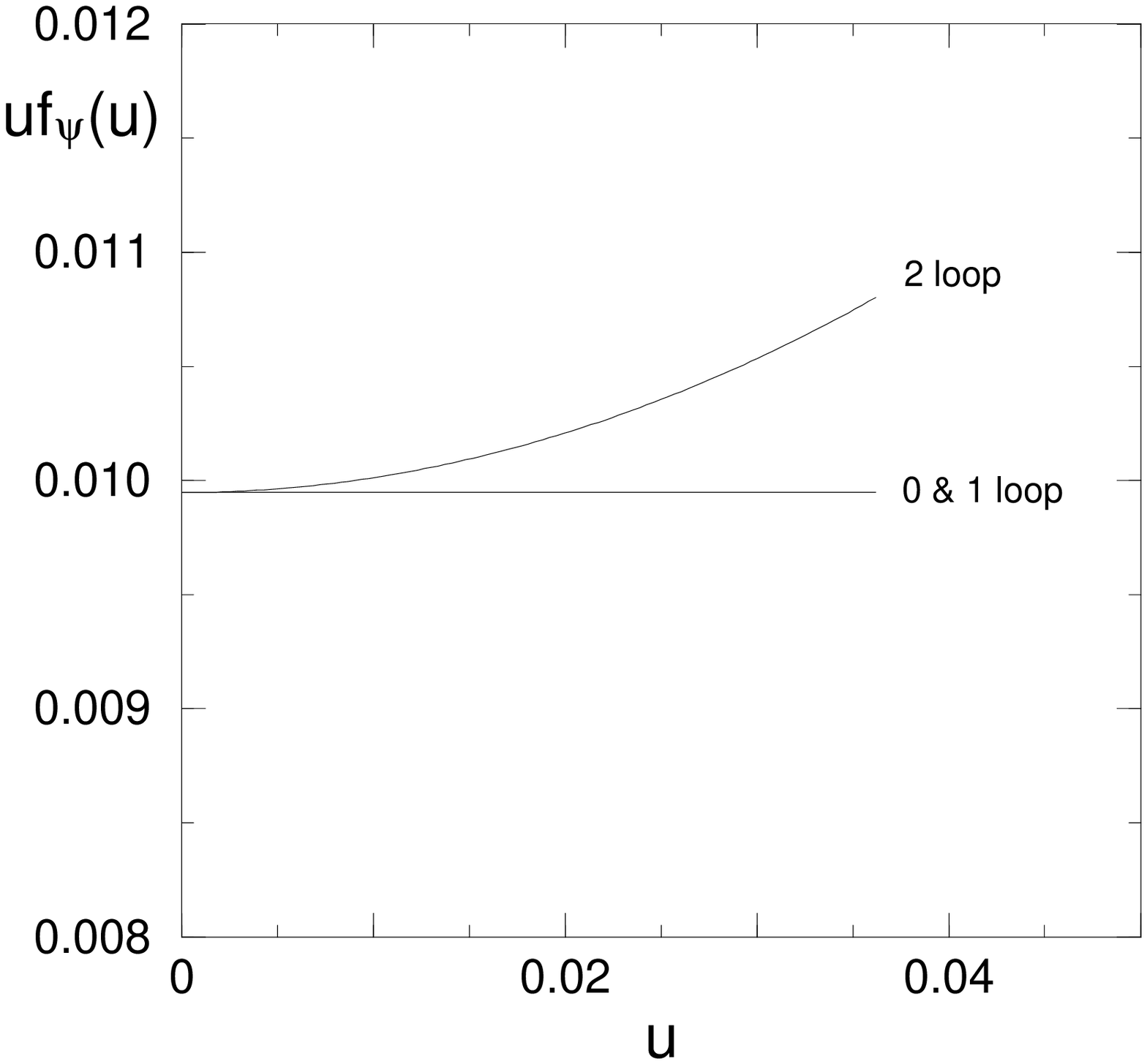,width=9cm}

        \vspace{5mm}
        \hspace*{2.7mm}\psfig{figure=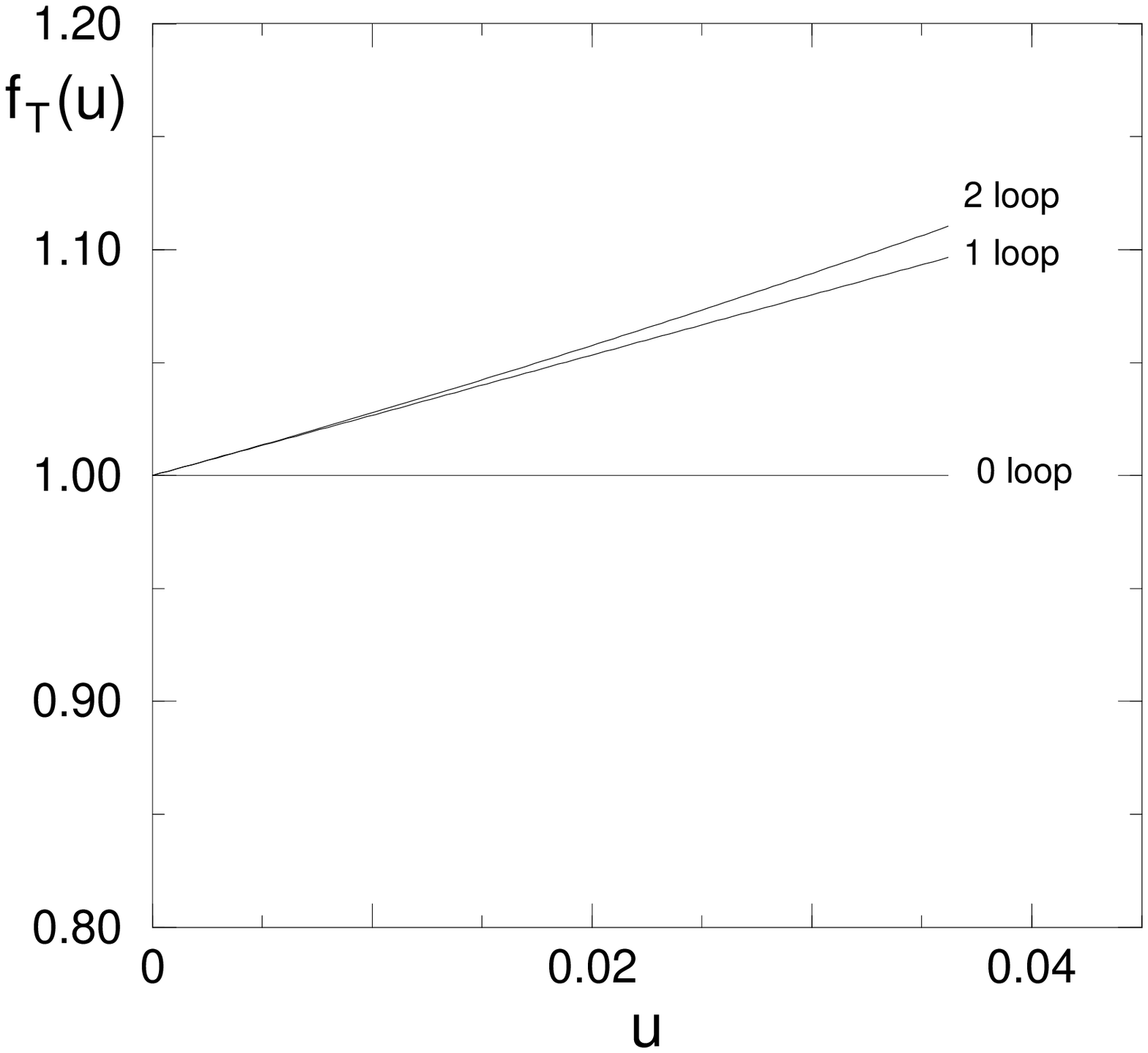,width=8.6cm}
      \end{center}

Fig.~1: Amplitude functions $uf_\psi(u)$ and $f_T(u)$ for the order
parameter and for the $k^2$ part of the inverse of the transverse 
susceptibility in zero-, one- and two-loop order vs the renormalized
coupling $u$. The curves terminate at the fixed point $u^* = 0.0362$.
>From Ref.~7.
\hspace*{\fill}

\newpage
While the two-loop correction to $f_T$ is 
remarkably small 
(about 1\%) at the fixed point $u^* = 0.0362$,
the two-loop contribution to $f_\psi(u)$ is about 10 \%, 
thus indicating a considerable uncertainty of low-order
perturbation theory. 
The uncertainty is even larger for the derivative of $f_\psi(u)$ at $u^*$
which implies a correspondingly large uncertainty of $a_\rho$ in (1) and of
the ensuing amplitude ratio $a^-_c/a_\rho$.
A similar uncertainty exists with regard to the
amplitude function $F_-(u)$ of the specific heat $C^-$ below $T_\lambda$ 
\cite{7,13} determining $A^-$ and $a^-_c$ in (3) as well as 
with regard to the amplitude function $G(u)$ of the
helicity modulus $\Upsilon$ \cite{7} which is proportional to $\rho_s$.

\vspace{1cm}

\section{New theory project}

In order to significantly reduce the uncertainty of the theoretical
predictions on the amplitude ratios (4) it has been proposed \cite{8} 
to perform
new higher-order field-theoretic RG calculations and Borel resummations
of various amplitude functions. In general, the results of such calculations
depend on the specific renormalization scheme employed. In view of
various applications, the method of the
minimally renormalized $\varphi^4$ field theory at fixed dimension
$d=3$ \cite{7,13,17} appears to be most appropriate.
The following quantities need to be calculated:
\begin{enumerate}
\item[(i)]
The additive renormalization constant $A(u,\varepsilon)$ of the
specific heat and the related RG function $B(u)$. At present, $A(u,\varepsilon)$
and $B(u)$ are known only in two-loop order \cite{17}. For a three-loop 
calculation within a different
renormalization scheme see \cite{20}.
\item[(ii)]
The amplitude function $f_\psi(u)$ of the square of the order parameter 
appearing in Eq. (6).
For $n=1$ it is known up to five-loop order and in Borel-resummed form
\cite{18} but for $n > 1$ only up to two-loop order \cite{7}.
\item[(iii)]
The amplitude function $F_-(u)$ of the specific heat below $T_\lambda$.
For $n=1$ it is known up to five-loop order and in Borel-resummed
form \cite{18} but for $n > 1$ only the two-loop result is known at present 
\cite{13}.
(Even for $n=1$ the five-loop result needs to be revised on the basis of
higher-order calculations for $B(u)$ according to item (i) above.)
\item[(iv)]
The amplitude function $F_+(u)$ of the specific heat above $T_\lambda$.
Previously this function has been computed in Borel-resummed form for
$n=1,2,3$ on the basis of six-loop results \cite{19}. 
But this calculation needs to be revised on the basis of higher-order
calculations for $B(u)$ according to item (i) above.
\item[({v})]
The amplitude function $f^-_\xi(u)$ of an appropriately defined correlation
length $\xi_-$ below $T_\lambda$.
This correlation length performs the task of absorbing logarithms in the
four-point coupling $u_0$ in the bare perturbation series in 
$d=3$ dimensions
\cite{7,13,18}. For $n=1$, $\xi_-$ can be taken to be the ordinary 
correlation length of three-dimensional Ising-like systems below $T_c$.
For a three-loop calculation of $\xi_-$ for $n=1$ within a different renormalization
scheme see \cite{21}.
\item[({vi})]
The amplitude function $f_T (u)$ [see Eq. (6)] of the $k^2$ part of the
inverse of the transverse correlation function $\chi_T(k)$. At present 
this function is known in two-loop order \cite{7}. 
Equivalently (and preferably), the amplitude function $G(u) = 4\pi f_\psi (u)
f_T (u)$ of the helicity modulus $\Upsilon$ \cite{7} should be calculated
in higher order. 
\end{enumerate}

Higher-order calculations of $G(u)$ or $f_T(u)$ would be more demanding
than those of $f_\psi$ and $F_-$ because they involve perturbation theory at 
finite wave number $k$.
In view of the smallness of the two-loop correction to $f_T(u)$ 
(see Fig. 1) it has been argued
\cite{7} that this low-order result for $f_T(u)$ can be considered as 
rather reliable, presumably within a few percent. In a first step of future calculations it will therefore be
sufficient to confine the project to the quantities (i) - (v) noted above.\\
\\

Once the functions $B(u), f_\psi(u), F_-(u), F_+(u)$ and $f^-_\xi(u)$ are known
at an accuracy of a few percent it will be possible to determine the amplitude
ratios $(4)$  at the same level of accuracy. It will be advantageous
to go beyond the asymptotic representations (1) - (3) and, instead, include
the entire Wegner series, i.e. higher-order terms $|t|^{n\Delta}$, 
$n = 2,3, \cdot \cdot \cdot$, {\it{without introducing new nonuniversal 
parameters into the analysis}}. For the strategy of such a 
{\it{nonlinear RG analysis}} see Refs. 6 and 17. This may imply 
that the phenomenological  analytic corrections $\sim k_1|t|$ and $ \sim E^\pm |t|$ in the
representations (1) - (3) are unnecessary or
will turn out  to be much smaller than determined  previously \cite{2,5,9}.\\
\\

The success of this theory project depends on the order of 
perturbation theory
up to which the calculations  can be carried out. 
Preliminary  considerations by Larin \cite{22} indicate that 
the amplitude functions 
$f_\psi(u), F_-(u)$ and $f^-_\xi$(u) can be determined  
up to four-loop order, and the RG functions $A(u,\varepsilon)$ and $B(u)$, 
on the basis of previous
work \cite{11}, up to five-loop order. Such results will 
have an important
impact on a future test of the fundamental RG prediction of critical-point 
universality.\\ 
\\

{\underline{\bf Acknowledgement}}\\
\\
I appreciate fruitful collaboration with S.S.C. Burnett and M. Str\"osser
on the work quoted in Refs. 7 and 15. I also thank S.A. Larin for
encouraging discussions on higher-order calculations.
Support by DARA and NASA is greatfully acknowledged.\\
\\

{\underline{\bf Note added}}\\
\\
Very recently a five-loop calculation of $A(u,\varepsilon)$ and
of $B(u)$ has been performed for general $n$ and a Borel resummation
of $B(u)$ has been carried out for $n = 1,2,3$ \cite{23,24}.
The Borel resummed function $B(u)$ differs from the two-loop 
result $B(u) = \frac{n}{2} + {\it{O}} (u^2)$ by less than 1 \% at 
the fixed point $u^*$. For $n = 2$ the result is $B(u^*) = 1.0053 \pm 
0.0022 $. Furthermore, a three-loop calculation of 
the amplitude functions $f_\psi(u)$ and $F_-(u)$
in three dimensions below $T_c$ has been carried out for general
$n \geq 1$ \cite{25,26}.

\end{document}